\documentclass[%
 reprint,
 amsmath,amssymb,
 aps,
]{revtex4-1}

\usepackage{graphicx}
\usepackage{dcolumn}
\usepackage{bm}
\usepackage{siunitx}
\usepackage{pdfcomment}

\begin{document}


\title{Oleoplaning droplets on lubricated surfaces}
\author{Dan Daniel}
\author{Jaakko V. I. Timonen}
\author{Ruoping Li}
\author{Joanna Aizenberg}%
 \email{jaiz@seas.harvard.edu}

\affiliation{%
  John A. Paulson School of Engineering and Applied Sciences, Harvard University, Cambridge, MA 02138}%
  
\maketitle

\textbf{
The ability to create materials with extreme liquid repellency has broad technological implications in areas ranging from drag reduction, to prevention of icing and biofouling \cite{bocquet2011smooth, quere2008wetting}. Recently, there has been much interest in using lubricated flat and nano-/micro-structured surfaces to achieve this property: any foreign droplet immiscible with the underlying lubricant layer was shown to slide off at a small tilt angle $<$ 2$^{\circ}$ \cite{lafuma2011slippery, wong2011bioinspired, grinthal2013mobile, kim2012liquid, epstein2012liquid, solomon2014drag, daniel2013lubricant}. This extreme non-wetting was hypothesized to arise from a thin lubricant overlayer film sandwiched between the droplet and solid substrate, but this has not been observed experimentally \cite{schellenberger2015direct, smith2013droplet}. Such demonstration is critical for understanding the mechanistic origin of liquid repellency on these surfaces and optimization of their function. Here, using confocal optical interferometry, we are able to visualize the intercalated film under both static and dynamic conditions. Based on the theoretical and experimental analysis of spreading coefficients and Hamaker constants, we rationalize the wetting/dewetting transitions of this film for 26 different material combinations for both flat and structured surfaces, and identify three film stability states. We further demonstrate how lubricant flow entrained by droplet motion can generate sufficient hydrodynamic force to lift the droplet over the solid substrate. The droplet is therefore oleoplaning---akin to tires hydroplaning on a wet road---with minimal dissipative force (down to 0.1 $\mu$N for a 1 $\mu$l droplet when measured experimentally) and no contact line pinning. The techniques and insights presented in this study will inform future work on the fundamentals of wetting for lubricated surfaces and enable their rational design.
}

On a flat substrate, a droplet experiences significant pinning forces at the three-phase contact line \cite{eral2013contact}. To overcome contact line pinning, the conventional approach is to design micro-/nano-structured surfaces that maintain a stable air layer within the structures, and hence minimize solid-liquid contact area (lotus-effect) \cite{quere2008wetting, reyssat2010dynamical}. A liquid droplet sitting on such a surface beads up into a ball with a high apparent contact angle, $\theta_{\text{app}} > 150^{\circ}$, and is able to roll off even at a small tilt, $\theta_{\text{tilt}}$, with negligible contact angle hysteresis, $\Delta \theta < 10^{\circ}$ (Cassie-Baxter state). There remain, however, obstacles to wide-spread adoption of such surfaces: the droplet can penetrate into the structures, displace the air layer and become highly pinned (Wenzel state) \cite{lafuma2003superhydrophobic, reyssat2006bouncing, tuteja2008robust}; exposure to fog can similarly defeat the liquid-repellency of lotus-effect surfaces \cite{dorrer2007condensation}; the micro-/nano-structures also typically lack mechanical robustness, though recently, there have been some progress in improving their mechanical durability \cite{verho2011mechanically, tian2016moving}. In any case, for lotus-effect surfaces, there are always some solid-liquid contact points, which serve as nucleation/attachment points for ice formation and fouling organisms \cite{sojoudi2016durable, genzer2006recent}. 

\begin{figure}
\includegraphics[scale=1.0]{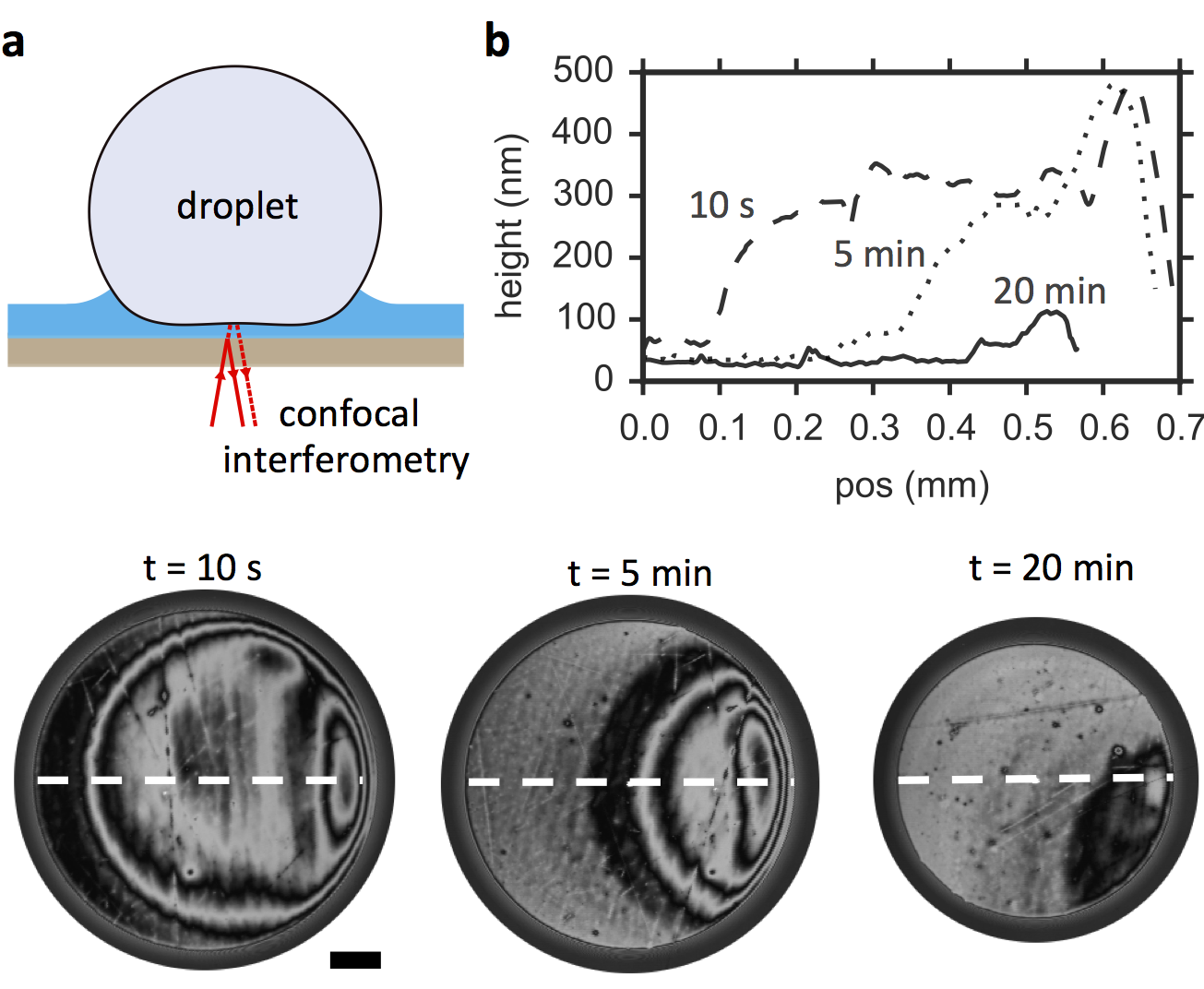} 
\caption{\label{fig:newton} Visualization of the lubricant film profile between the droplet and the solid using optical interferometry. a, Schematic of the confocal optical interferometry set-up. b, Evolution of the silicone oil film thickness between a water droplet and a flat polymethylpentene (PMP) substrate with time, along the dashed lines drawn on the images shown below. Droplet size is decreasing due to evaporation. Scale bar is 0.1 mm.
}  
\end{figure}

\begin{figure*}
\includegraphics[scale=1]{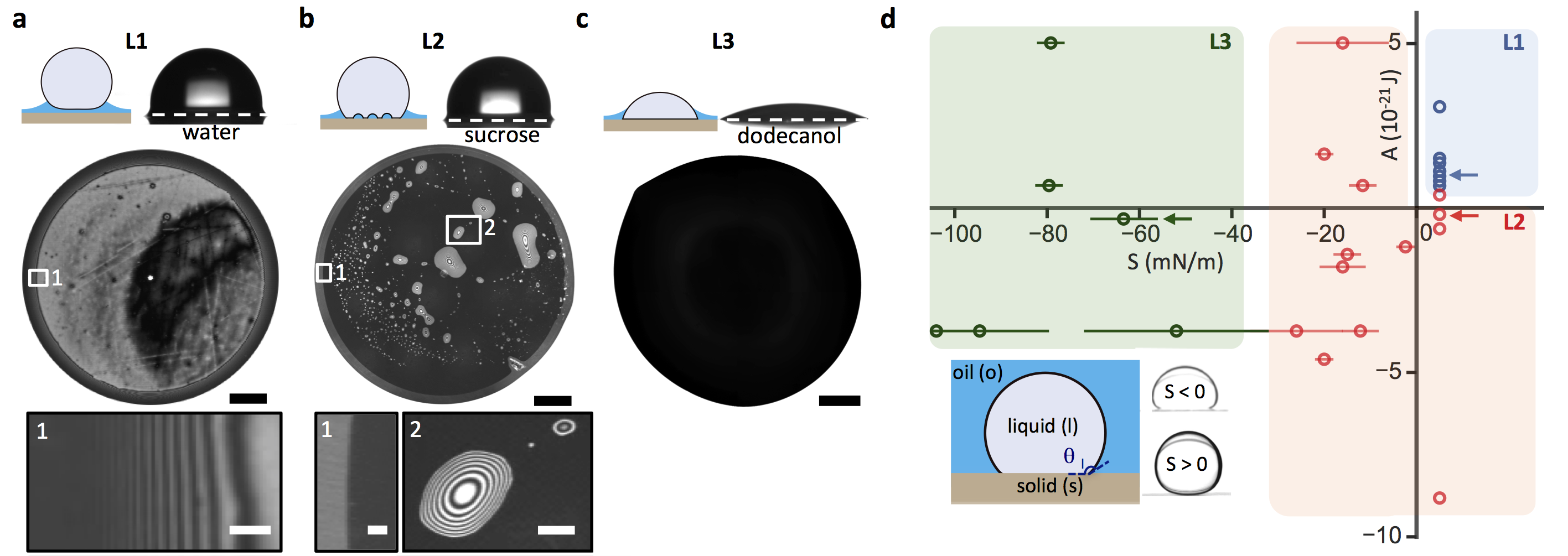} 
\caption{\label{fig:static_flat} Equilibrium lubrication states L1--3 of an oil film sandwiched between a droplet and a flat solid surface. a, L1: stable lubricant film (silicone oil) beneath a water droplet (0.1 mm scale bar, $\theta_{\text{app}}$ = 90$^{\circ}$), with inset 1 (10 $\mu$m scale bar) showing the absence of a contact line. b, L2: lubricant film is unstable beneath 60 wt $\%$ sucrose solution and form discrete pockets (0.3 mm scale bar, $\theta_{\text{app}}$ = 90$^{\circ}$). Inset 1 (15 $\mu$m scale bar) shows the three-phase contact line, while inset 2 (40 $\mu$m scale bar) is a zoomed-in image of the lubricant pockets. c, L3: lubricant is completely displaced from beneath a dodecanol droplet (0.1 mm scale bar, $\theta_{\text{app}}$ = 23$^{\circ}$). Solid substrate is PMP for a--c. d, Phase diagram of the lubrication states L1--3 (shaded blue, red and green, respectively) determined for 26 different combinations of substrate, lubricant oil, liquid and surface treatment. Data points corresponding to droplets in a--c are marked with arrows. Error bars in $S$ comes from uncertainty in measuring $\theta_{l}$. See Supplementary table S1 for the data used to generate the phase diagram.  
}  
\end{figure*}

In contrast, for a lubricated surface, it has been postulated that there is a continuous, intercalated lubricant film with no contact between droplet and solid, giving it its liquid-repellent, anti-icing and anti-fouling properties \cite{grinthal2013mobile, kim2012liquid, epstein2012liquid}. Previous attempts to visualize this film, however, have yielded inconclusive results, because the technique used, namely confocal fluorescence microscopy, lacks the resolution to observe sub-micron-thick lubricant film \cite{smith2013droplet, schellenberger2015direct}. Moreover, fluorescent dyes added to a liquid and/or lubricant oil may act as surfactants, and change their wetting properties. Here, we made use of thin-film interference effect instead, without the addition of a dye to the droplet or lubricant (Fig.~\ref{fig:newton}a). Thin-film interference  has been previously used together with conventional wide-field imaging optics to study thin air film beneath a droplet as it impacts a solid substrate \cite{de2015air}. However, on lubricated surfaces, the weak refractive index contrast between solid, lubricant and droplet leads to a much weaker reflection signal that is overwhelmed by stray light. To overcome this problem, we raster scanned the surface with focused monochromatic light of wavelength, $\lambda = 405$ nm, and captured the reflected light through the pinhole of a confocal microscope; as a result, only light from the focal plane, i.e. reflected off the interface of interest, is able to reach the photomultiplier tube of the microscope. In the presence of a thin lubricant film, the light reflected off the solid-lubricant and lubricant-droplet interfaces will then interfere with one another constructively or destructively to give bright or dark fringes, respectively. Between adjacent bright and dark fringes, there is a difference in film thickness, $|\Delta h| = \lambda/4n_{o} \approx 100$ nm, where $n_{o}$ is the refractive index of the lubricant oil. Using this technique, the lubricant film profile can be deduced from the reflected light intensity with nanometric resolution (Supplementary Fig.~S1).
 
First, we studied the equilibrium state of the lubricating thin film under a static (non-moving) droplet of another liquid, and observed three different distinct lubrication states: L1-3. The first lubrication state (L1) corresponds to a stable lubricant film and was observed in the case of silicone oil of viscosity, $\eta = 10$ cP, sandwiched between a 0.5 $\mu$l water droplet and a flat, transparent polymethylpentene (PMP) substrate (Fig.~\ref{fig:newton}). Without the droplet, the initial lubricant thickness, $h_{\text{init}}$, as measured using white light interferometry, was about 3 $\mu$m. Within 10 s of placing the water droplet, however, the average intercalated film thickness, $h$, decreased to approximately 300 nm. After 20 min, the film thickness stabilized to its equilibrium value of $h_{\infty} = 30 \pm 5$ nm (Fig.~\ref{fig:newton}b). Thus, we confirm that the hypothesized stable, continuous lubricating thin film can indeed exist (Fig.~\ref{fig:static_flat}a). 

However, this film can easily be destabilized, for example, by replacing the water droplet with 60 wt$\%$ aqueous sucrose solution (lubrication state L2). Silicone oil dewets under the sucrose solution and forms small lubricant pockets that are stable in time (Fig.~\ref{fig:static_flat}b). In addition, a clear contact line becomes visible at the droplet base for L2 (Fig.~\ref{fig:static_flat}b-1), which in contrast is missing in L1 (Fig.~\ref{fig:static_flat}a-1) We note that despite the very different lubricant behavior at the microscopic scale, droplets in the two lubrication states L1 and L2 still exhibit the same apparent macroscopic contact angle $\theta_{\text{app}}$ = 90$^{\circ}$ and are thus practically impossible to distinguish using conventional contact angle goniometry. However, as we will show, droplets nevertheless experience very different dissipative forces depending on their lubrication state, L1 or L2. In the third lubrication state (L3), lubricant is completely displaced under the droplet. This was the case for a dodecanol droplet on the same PMP substrate lubricated with silicone oil. The dodecanol droplet was irregularly shaped, highly pinned to the substrate with contact angle, $\theta_{\text{app}}$ = 23$^{\circ}$ (Fig.~\ref{fig:static_flat}c). In addition to these examples, we looked at 23 other combinations of solid, lubricant oil, liquid and surface treatment, and found that all of them could be classified into these three lubrication categories (Supplementary table S1).

We rationalize the lubrication states for different material combinations (solid, lubricant and liquid droplet) and surface treatments to originate from a combination of 1) interfacial tensions effect and 2) short-ranged van der Waals' interaction (Fig.~\ref{fig:static_flat}d). The former is represented by the spreading constant, $S = \gamma_{ls} - (\gamma_{lo} + \gamma_{os})$, where $\gamma_{ls}$, $\gamma_{lo}$ and $\gamma_{os}$ are the liquid-solid, liquid-lubricant oil, lubricant oil-solid interfacial tensions, and the latter by the Hamaker constant, $A$. For $S < 0$, $S$ can be determined by measuring the contact angle of the droplet, $\theta_{l} < 180^{\circ}$, while submerged inside the lubricant oil and using the relation $S = -\gamma_{lo}(\cos \theta_{l} + 1)$ \cite{de2013capillarity}. When $\theta_{l} = 180^{\circ}$, $S \geq 0$. Earlier work has pointed out that for a lubricant film to be stable, $S > 0$ \cite{lafuma2011slippery, smith2013droplet}. This, however, is not a sufficient condition, since $S > 0$ is the stability condition for a micron-thick film in the absence of other external destabilizing forces. A lubricant film beneath a droplet of radius $R$, on the other hand, is being continuously squeezed out with pressure, $P \sim \gamma/R$ (Supplementary Fig.~S2), and can only be stabilized, if the disjoining pressure due to van der Waals' interaction, $\Pi(h) = A/(6 \pi h^{3})$, in the lubricant film is positive, i.e. if $A > 0$ \cite{de2013capillarity, brochard1991spreading}. The coefficient $A$ can be estimated by using non-retarded Hamaker constant in Lifshitz theory:
\begin{equation} \label{eq:Hamaker2}
\begin{split}
A &= \frac{3}{4}K_{B}T \left ( \frac{\epsilon_{o}-\epsilon_{l}}{\epsilon_{l}+\epsilon_{o}} \right)\left ( \frac{\epsilon_{s}-\epsilon_{o}}{\epsilon_{s}+\epsilon_{o}} \right) + \\
&\frac{3\pi \hbar \nu_{e}}{4\sqrt{2}} \frac{(n_{o}^{2}-n_{l}^{2}) (n_{s}^{2}-n_{o}^{2})}{\sqrt{(n_{l}^{2}+n_{o}^{2})(n_{s}^{2}+n_{o}^{2})} [\sqrt{(n_{l}^{2}+n_{o}^{2})} + \sqrt{(n_{s}^{2}+n_{o}^{2})}]}\text{,}
\end{split}
\end{equation}
where $\nu_{e} \approx 4 \times 10^{15} s^{-1}$ is the plasma frequency of free electron gas, while $\epsilon_{l/o/s}$ and $n_{l/o/s}$ are the dielectric constants and refractive indices of the liquid droplet, oil lubricant and solid substrate, respectively \cite{israelachvili2011intermolecular}. 

Hence, for a stable, continuous lubricant layer, i.e. lubrication state L1, two criteria must be met: $S > 0$ and $A > 0$. For typical material combinations, $\gamma \sim 50$ mN/m and $|A| \sim 10^{-21}$ J, and therefore at equilibrium, when $\Pi = P$, the equilibrium thickness $h_{\infty} \sim (RA/\gamma)^{1/3}$ is about tens of nm, in agreement with our experimental observations (Fig.~\ref{fig:newton}). If the two criteria are not met, the lubricant film will dewet either partially (L2) or completely (L3) in the extreme case when $S \lesssim -30$ mN/m, as summarized in the phase diagram of Fig.~\ref{fig:static_flat}d. Note that, in general $A > 0$ when the optical properties of the lubricant oil, in particular its refractive index, are intermediate of those of the liquid droplet and the solid substrate, i.e. $n_{s} > n_{o}> n_{l}$ (equation (\ref{eq:Hamaker2})) \cite{israelachvili2011intermolecular, de2013capillarity}. This is the case for stable silicone oil, $n_{o} = 1.41$, sandwiched between a water droplet, $n_{l} = 1.33$, and PMP substrate, $n_{s} = 1.46$ (Fig.~\ref{fig:static_flat}a), but not when the water droplet is replaced with a 60 wt $\%$ sucrose solution, $n_{l}=1.44$ (Fig.~\ref{fig:static_flat} b). Hence, the silicone oil dewets into pockets under the sucrose droplet, even though $S > 0$. See Supplementary Figs~S3 and S4 for details. 

\begin{figure*}
\includegraphics[scale=1.0]{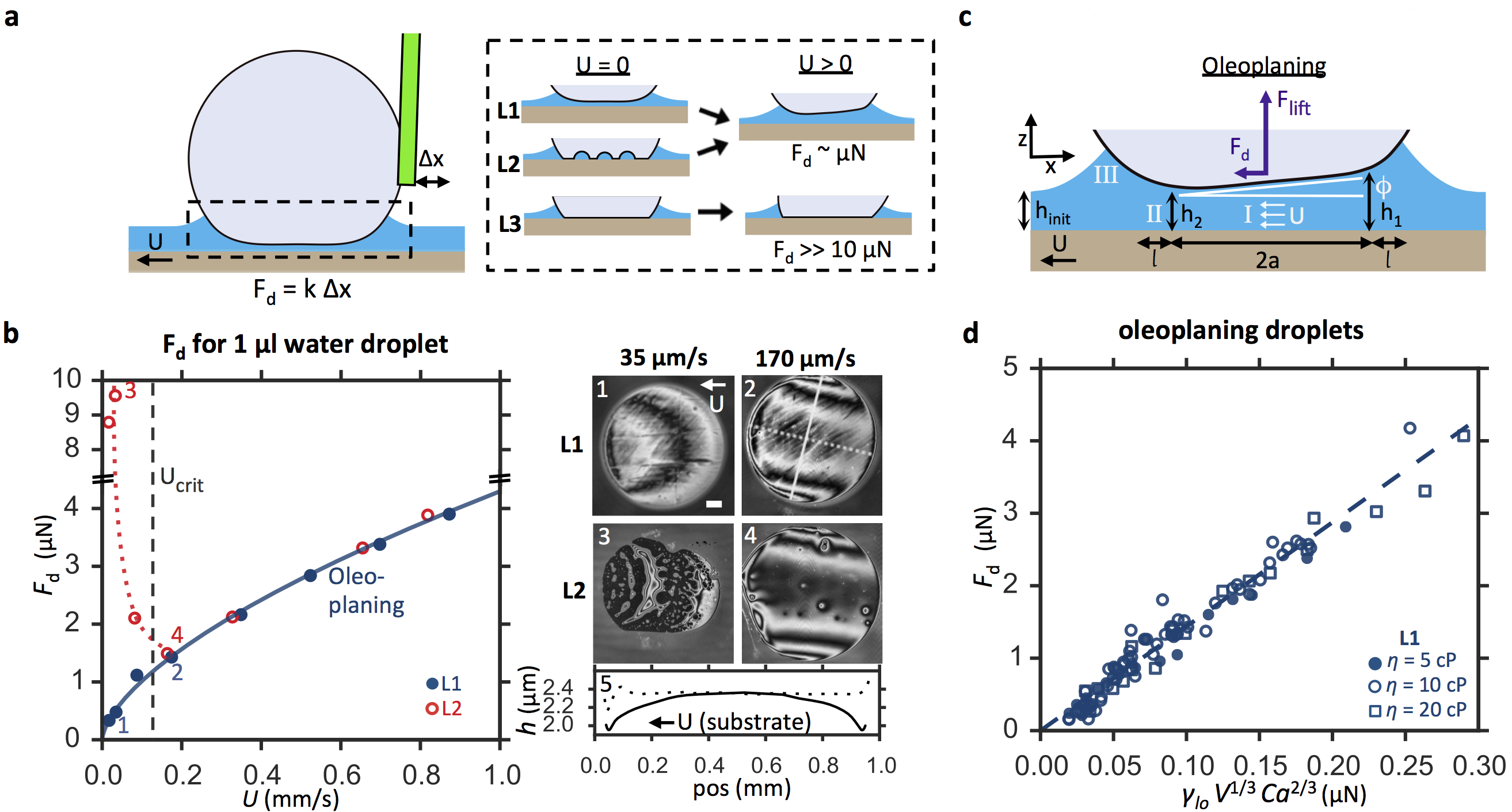} 
\caption{\label{fig:dynamic_flat} Dynamic lubrication states and dissipative force, $F_{d}$, acting on a droplet moving on flat, lubricated surfaces. a, Schematic of the cantilever force sensor to measure $F_{d}$, which depends on the lubrication state and lubricant dynamics beneath the droplet. b, Plot on the left shows $F_{d}$ acting on 1 $\mu$l water droplet moving on lubricated substrate ($\eta$ = 30 cP) for L1 and L2. Each data point is repeated at least 3 times, with a standard deviation, $\Delta F_{d} < $  0.3 $\mu$N. Solid line is the prediction of equation (\ref{eq:viscous}), while the dotted line is a guide to the eye. Insets 1--4 on the right (0.1 mm scale bar) depict interferometry images that show the stability of the intercalated film around $U_{\text{crit}}$ for L1 and L2, while inset 5 is the lubricant film profile along solid and dashed lines in inset 2. Data points corresponding to insets 1-4 are similarly marked on the $F_{d}$ vs $U$ plot. See Supplementary section S2 for details on sample preparation for L1 and L2: glass hydrophobized with $\sim$ 200 nm teflon-like coating and vapor-phase silanization, respectively. c,  Schematic of droplet oleoplaning on lubricated surface, showing the asymmetry of the lubricant film profile and the parameters used in equation (\ref{eq:lift}). d, Experimental data of $F_{d}$ measured for $V$ = 1--5 $\mu$l of water droplets oleoplaning on PMP surface lubricated with silicone oil of different viscosities, $\eta$ = 5--20 cP, follows a scaling law given by equation (\ref{eq:viscous}) (dashed line). Errors in $F_{d}$ are less than 0.3 $\mu$N.}  
\end{figure*} 

To quantify the `slipperiness' of lubricated surfaces, previous work typically reports the $\theta_{\text{tilt}}$ of a probe liquid and/or its apparent $\Delta \theta$, as measured optically using a contact angle goniometer. However, this technique cannot distinguish between L1 and L2 states (Fig.~\ref{fig:static_flat}a,b and Supplementary movie S1). To characterize the differences between these two states, we built a customized cantilever force sensor and measured the dissipative force, $F_{d}$, acting on a moving droplet with a sensitivity of about 0.05 $\mu$N (Fig.~\ref{fig:dynamic_flat}a). Briefly, the droplet was attached by its own capillarity to an acrylic tube with inner and outer radii of 0.29 and 0.36 mm, respectively, while the substrate was moved at a controlled speed $U$ in the range of 0.01-1 mm/s (See Supplementary Fig.~S5 for details of the set-up). $F_{d}$ can then be inferred from the deflection of the capillary tube $\Delta x$, since $F_{d} = k \Delta x$, where $k = 6 \times 10^{-3}$ N/m. 

We observed that $F_{d}$ increases monotonically with $U$ for a 1 $\mu$L water droplet in L1 lubrication state, with $F_{d} \propto U^{2/3}$, and is independent of the initial lubricant height, $h_{\text{init}}$ (Fig.~\ref{fig:dynamic_flat}b). Furthermore, we observed no pinning of the droplet even at lowest experimentally realizable velocity of 10 $\mu$m/s and $F_{d} \rightarrow 0$ as $U \rightarrow 0$. We attribute this behavior to the stable lubricant layer that prevents pinning and gives rise to velocity-dependent, viscous dissipative force. In contrast, a 1 $\mu$L water droplet in L2 lubrication state showed clear contact line pinning with $F_{d} = 9 \pm 1$ $\mu$N as $U \rightarrow 0$. However, a sudden decrease in $F_{d}$ was observed above a critical velocity $U_{\text{crit}} = 150$ $\mu$m/s. We attribute this transition to a velocity-dependent lift force generated by droplet motion, which stabilizes the intercalated lubricant film (Fig.~\ref{fig:dynamic_flat}b-3,4, c.f. Fig.~\ref{fig:dynamic_flat}b-1,2). The droplet is, therefore, oleoplaning, akin to tires hydroplaning on a wet road. As long as the droplet is oleoplaning, $F_{d} (U)$ is indistinguishable between L1 and L2 for the same droplet (1 $\mu$l water droplet) and lubricant (perfluorinated oil, $\eta = 30$ cP). Not surprisingly, droplets in L3 are highly pinned, with $F_{d} \gg$ 10 $\mu$N.
 
The lift force $F_{\text{lift}}$ depends on the shape of the lubricant film under the droplet base. For example, during oleoplaning of a 1 $\mu$l water droplet moving at $U$ = 170 $\mu$m/s, the film takes the shape of a saddle, with a maximum thickness at the advancing front, a local minimum at the receding front, and global minima at the two side rims (Fig.~\ref{fig:dynamic_flat}b-2,5). We can approximate this geometry as a slightly-tilted plane with length $2a$, and a small tilt angle, $\phi = (h_{1}-h_{2})/2a$ (Fig.~\ref{fig:dynamic_flat}c), in which case: 
\begin{equation} \label{eq:lift}
\begin{split}
F_{\text{lift}} &= \frac{6 \eta a U}{\phi^{2}}\bigg[\ln \frac{2+\delta}{2-\delta}-\delta \bigg] \approx \frac{\eta a U}{\phi^{2}}\frac {\delta^{3}}{2}\text{,}
\end{split}
\end{equation}
where $h_{1},h_{2} = 2.4, 2.3 \pm 0.1$ $\mu$m are the thicknesses at the advancing and receding fronts, respectively, and $\delta = 2 (h_{\text{1}}-h_{\text{2}})/(h_{\text{1}}+h_{\text{2}})$ \cite{batchelor2000introduction, martin1998dewetting}. For steady droplet motion, the resultant $F_{\text{lift}} \approx 150$ $\mu$N balances the capillary force pulling down the droplet, $F_{\gamma} \sim \gamma R \approx 100$ $\mu$N. This approximation, while simple, captures the essential point that the asymmetry of the lubricant film profile in the direction of the droplet motion is crucial in generating a lift force (Supplementary Fig.~S6).

The functional form of the dissipative force $F_{d}$ can be derived as follows. First we note that the lubricant is more viscous than the droplet in our experiments, and thus the droplet rolls while oleoplaning. This can be confirmed by seeding the droplet with tracer particles (See Supplementary Fig.~S7). Beneath the droplet (region I, Fig.~\ref{fig:dynamic_flat}c), there is therefore plug flow of the lubricant (in the reference frame of the stationary droplet), i.e. $\nabla_{z} U = 0$, and hence minimal viscous dissipation. Most of the viscous dissipation occurs instead at the small transition region II of size $l$ at the rim of region I, which experiences a gradient of Laplace pressure, $\gamma_{lo}/Rl$, and a typical viscous stress, $\eta U/h$. The thickness of the intercalated film, $h$, can be predicted by balancing $\nabla P$ and $\eta \nabla^{2} U$ in region II, i.e. $\gamma_{lo}/Rl \sim \eta U/h^{2}$, and matching the curvature in this transition region, $\partial^{2}h/\partial x^{2} \sim h/l^{2}$, with that of the main droplet, $1/R$, i.e.  $h/l^{2} \sim 1/R$. This gives $h \sim R Ca^{2/3}$ and $l \sim R Ca^{1/3}$, where $Ca = \eta U/\gamma_{lo}$ is the capillary number \cite{bretherton1961motion, cantat2013liquid, de2013capillarity}. For a typical $Ca = 10^{-4}$, $h \ll l \ll R$, and hence justifies the use of lubrication approximation in this analysis. The dissipative force, $F_{d}$, can then be estimated by integrating the viscous stress, $\eta U/h$, over the area $2 \pi a l \approx 2 \pi Rl \approx 2 \pi (3V/2\pi)^{1/3} l$: 
\begin{equation} \label{eq:viscous}
\begin{split}
F_{d} \approx (\eta U/h) 2\pi al \approx \beta_{1} \gamma_{lo} V^{1/3} Ca^{2/3}\text{,}    
\end{split}
\end{equation}
where $V$ is the droplet volume and $\beta_{1} = 3^{1/3} (2\pi)^{2/3}$ is a non-dimensional prefactor. We confirmed this scaling of $F_d$ experimentally for $V$ = 1-5 $\mu$l water droplets moving at $U = 0.01$-5 mm/s on PMP substrate lubricated with silicone oil of viscosity $\eta$ = 5-20 cP (Fig.~\ref{fig:dynamic_flat}d). Experimentally, the observed prefactor, $\beta_{1, \text{exp}} = 14$, is slightly larger than the predicted value, $3^{1/3} (2\pi)^{2/3} \approx 5$, due to the approximations made. We also note that the predicted scaling of $F_{d}$ is independent of the initial lubricant film thickness, $h_{\text{init}} \neq h$, or the size of the wetting ridge (region III, Fig.~\ref{fig:dynamic_flat}c), which we also observed experimentally (Supplementary Fig. S8).

\begin{figure}
\includegraphics[scale=1.0]{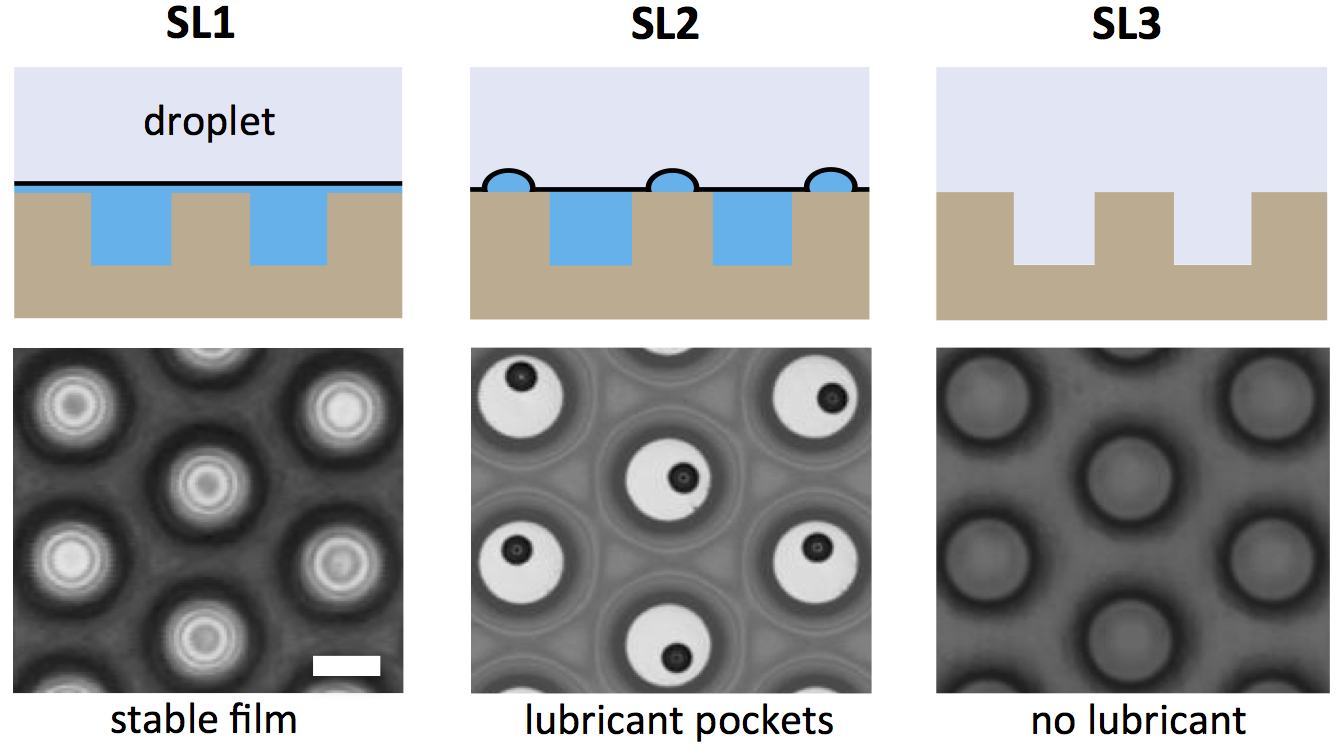} 
\caption{\label{fig:posts_lubricant} Lubrication states for lubricant-infused surfaces bearing a hexagonal array of microposts with diameter $D$ = 26 $\mu$m, pitch $p$ = 50 $\mu$m and height $h_{\text{o}}$ = 30 $\mu$m. The liquid droplet is water, while the solid substrate is made from UV-cured polymer (NOA 61, Norland), lubricated with perfluorinated oil. Different wetting states are achieved by different surface treatment: SL1, 200 nm teflon-like coating; SL2, vapor-phase silanization with perfluorosilane; SL3, no surface treatment. See Supplementary section S2 for details on sample preparation. Scale bar is 25 $\mu$m.}
\end{figure}

Many lubricated surfaces reported in literature have nano/microstructures on them to hold the lubricant oil in place for durability \cite{lafuma2011slippery, smith2013droplet,wong2011bioinspired, kim2012liquid, epstein2012liquid, solomon2014drag}. The analysis presented above for flat surfaces can similarly be applied to structured surfaces. For a solid substrate with a hexagonal array of microposts, we observed three lubrication states SL1-3 for different solid-lubricant-droplet combinations (Fig.~\ref{fig:posts_lubricant}): there can be a stable, intercalated lubricant film (SL1), micron-sized lubricant pockets on top of posts (SL2), or complete displacement of lubricant under static conditions (See Supplementary section S8, and Supplementary movies S2 and S3). 

\begin{figure*}
\includegraphics[scale=1.0]{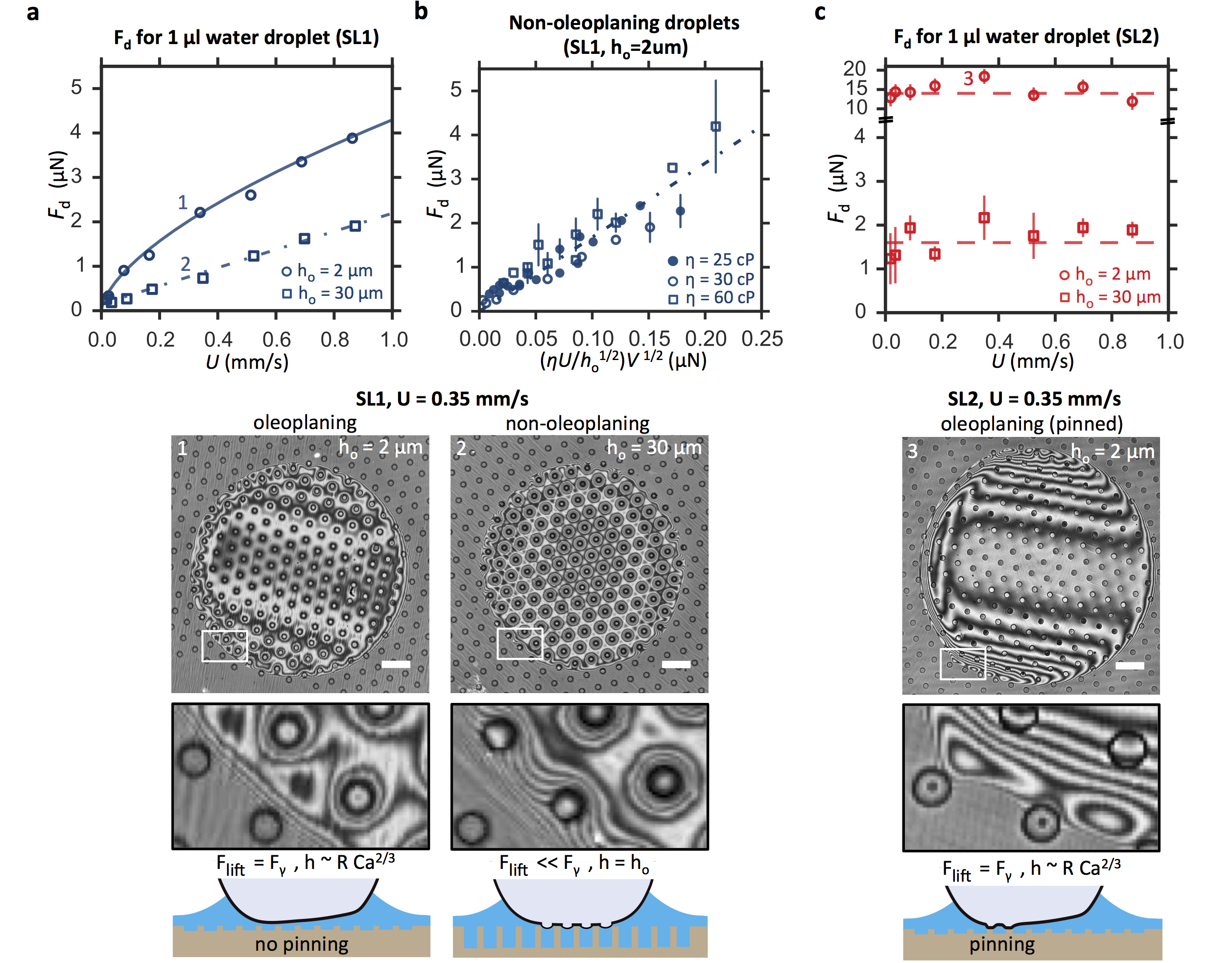} 
\caption{\label{fig:posts} Dynamic lubrication states and dissipative force, $F_{d}$, acting on droplet moving on micro-structured, lubricant-infused surfaces. The surface is decorated with a hexagonal array of posts, with diameter $D$ = 16 $\mu$m, pitch $p$ = 50 $\mu$m and heights $h_{\text{o}}$ = 2 and 30 $\mu$m. a, $F_{d}$ acting on a 1 $\mu$l water droplet in lubrication state SL1 for $h_{\text{o}}$ = 2 and 30 $\mu$m. The surface has a teflon-like coating and is lubricated with perfluorinated oil of $\eta$ = 30 cP. Depending on the micropost heights, the droplet will either oleoplane or not, resulting in dissipative force which scales with $U^{2/3}$ (solid line) or $U$ (dash-dot line), i.e. obeying equations (\ref{eq:viscous}) and (\ref{eq:viscous_posts}) respectively. b, Experimentally, equation (\ref{eq:viscous_posts}) is well obeyed for non-oleoplaning water droplets of volume $V$ = 1--3 $\mu$l moving on microposts surface of $h=30$ $\mu$m lubricated with perfluorinated oil of different viscosities $\eta$ = 25--60 cP. c, In contrast, for lubrication state SL2, $F_{d}$ is indepedent of $U$ for both oleoplaning and non-oleoplaning water droplet moving on micropost surfaces with $h_{\text{o}}$ = 2 and 30 $\mu$m respectively. The lubricant here is also perfluorinated oil of $\eta$ = 30 cP. Insets below are interferometry images depicting the dynamic lubrication states for oleoplaning and non-oleoplaning droplets for selected data points marked 1--3 on a and c. Scale bars are 0.1 mm. For a--c, each data point is repeated at least 3 times, with a standard deviation, $\Delta F_{d} <$ 0.3 $\mu$N, unless otherwise indicated by the error bars.}  
\end{figure*}

Even though the microscopic wetting behavior is analogous to lubrication states L1-3 on flat surfaces, the structured wetting states SL1-3 exhibit qualitatively different dissipation as the post height increases. For example, on a surface with low-aspect-ratio microposts of diameter $D$ = 16 $\mu$m, pitch $p$ = 50 $\mu$m and height $h_{\text{o}}$ = 2 $\mu$m, a 1 $\mu$l water droplet in SL1 state oleoplanes and experiences the same $F_{d} \propto U^{2/3}$ as its counterpart on a flat surface (solid lines in Fig.~\ref{fig:posts}a and Fig.~\ref{fig:dynamic_flat}b). However, increasing $h_{\text{o}}$ from 2 to 30 $\mu$m (while keeping $D$ and $p$ constant) led to different scaling of the force with velocity $F_{d} \propto U$ (dash-dot line in Fig.~\ref{fig:posts}a). This is because lubricant flow can no longer generate sufficient lift: $F_\text{lift} \ll F_{\gamma}$, since $F_\text{lift} \propto 1/h^{3}$ (equation (\ref{eq:lift})). Thus, the water droplet does not oleoplane but instead rolls on top of posts lubricated with nanometric film thickness, i.e. $h = h_{\text{o}}$ (insets 1 and 2, Fig.~\ref{fig:posts}). See Supplementary movies S4 and S5. Assuming further that, $h_{o}/l^{2} \sim 1/R$, $F_{d}$ is now given instead by:   
\begin{equation} \label{eq:viscous_posts}
\begin{split}
 F_{d} \approx (\eta U/h_{\text{o}}) 2 \pi al 
 \approx \beta_{2}(\eta U/h_{\text{o}}^{1/2})V^{1/2}\text{,}
\end{split}
\end{equation}
where $\beta_{2} =  (6\pi)^{1/2}$ is a non-dimensional pre-factor. This scaling was confirmed experimentally for $V$ = 0.5--3 $\mu$l water droplets moving at $U$ = 0.01--1 mm/s on surfaces with $h_{\text{o}}$ = 30 $\mu$m and different perfluorinated oils with different viscosities $\eta$ = 25--60 cP (Fig.~\ref{fig:posts}b). The experimentally observed pre-factor, $\beta_{2, \text{exp}} = 17$, is again slightly larger than the predicted value, $(6\pi)^{1/2} \approx 4$.

For a flat surface in L2, the intercalated film can be stabilized with $U > U_{\text{crit}} = 150$ $\mu$m/s. In contrast, for structured surfaces in lubrication state SL2, the microposts act as pinning defects, and $F_{d}$ is dominated by contact line pinning for all experimentally tested velocities up to $U$ = 1 mm/s (Fig.~\ref{fig:posts}c). Pinning dominates the dissipation even when the advancing front of the droplet is actually in oleoplaning state (inset 3, Fig.~\ref{fig:posts}). Energy is dissipated due to distortion of the contact line as the droplet retracts from one post to the next. We can estimate $F_{d}$ by assuming that the force due to each post $\sim \gamma D$, and hence $F_{d} \sim (2a/p) \gamma D$ = 8 $\mu$N. This is in reasonable agreement with experimentally observed $F_{d}$ = $15 \pm 5$ $\mu$N and $1.8 \pm 0.5$ $\mu$N for microposts with $h_{\text{o}}$ = 2 and 30 $\mu$m. We attribute the difference to the exact details of contact line distortion and number of posts to which the droplet was pinned. 

In summary, we have observed and explained the main static and dynamic wetting states on lubricated surfaces. A stable nanometric lubricant film can exist between a solid surface and a stationary liquid droplet when the film is stabilized by short-ranged van der Waals' interaction. However, even a lubricant film that dewets under a non-moving droplet can be dynamically stabilized into a continuous film by droplet motion that generates a lift force which pushes the droplet away from the surface into an oleoplaning state. We also pointed out that the conventional paradigm of characterizing the liquid-repellency and wetting-states of a lubricated surface by the contact angle, tilt angle or the contact angle hysteresis of a probe liquid is insufficient: droplets can exhibit the same tilt angle or contact angle hysteresis, but exhibit qualitatively different lubrication states and droplet mobility. A much more informative parameter is the dissipative force acting on the moving droplet, $F_{d}$, as a function of its speed, $U$, which can be dominated by either viscous dissipation or contact line pinning, depending on the absence or presence of the solid-droplet contact. Finally, our results have wide implications beyond motion of droplets on lubricated surfaces: stabilization of the lubricant thin film is crucial, for example, to prevent adhesion of ice, biomolecules and living micro-organisms on lubricated surfaces.

\textbf{Acknowledgements}. We thank Dr Kyoo-Chul Park and Dr C. Nadir Kaplan for careful reading of the manuscript. The work was supported partially by the ONR MURI Award No. N00014-12-1-0875 and by the Advanced Research Projects Agency-Energy (ARPA-E), U.S. Department of Energy, under Award Number DE-AR0000326. We acknowledge the use of the facilities at the Harvard Center for Nanoscale Systems supported by the NSF under Award No. ECS-0335765.

\providecommand{\noopsort}[1]{}\providecommand{\singleletter}[1]{#1}%

\end{document}